\documentclass[journal]{IEEEtran}

\usepackage{amsthm}
\usepackage{float}
\usepackage{cite}										
\usepackage{amsmath}									
\usepackage{graphicx}
\usepackage{epstopdf}
\usepackage{float}
\usepackage{bm,upgreek}
\usepackage{amsfonts}
\usepackage{algorithm,algpseudocode}
\usepackage{enumitem}
\usepackage{mathtools}
\usepackage{multirow}
\usepackage{booktabs}
\usepackage[subnum]{cases}
\usepackage{setspace}
\usepackage{color}
\usepackage{microtype}
\usepackage[caption=false,labelformat=simple]{subfig}
\usepackage{tikz}

\usetikzlibrary{positioning,shapes,arrows}
\usepackage{pgfplots}
\pgfplotsset{compat=1.5}
\usepackage{pgfplotstable}
\usepgfplotslibrary{statistics}
\pgfplotsset{grid style={dotted,gray}}
\usepackage[keeplastbox]{flushend} 
\newtheorem{example}{Example}

\allowdisplaybreaks
\emergencystretch=\maxdimen
\ifCLASSINFOpdf
\else
\fi
\pgfplotsset{legend image with text/.style={legend image code/.code={%
\node[anchor=west, align=right] at (0.0cm,0cm) {#1};}},}

\pgfplotsset{
    box plot/.style={
        /pgfplots/.cd,
        black,
        only marks,
        mark=-,
        mark size=\pgfkeysvalueof{/pgfplots/box plot width},
        /pgfplots/error bars/y dir=plus,
        /pgfplots/error bars/y explicit,
        /pgfplots/table/x index=\pgfkeysvalueof{/pgfplots/box plot x index},
    },
    box plot box/.style={
        /pgfplots/error bars/draw error bar/.code 2 args={%
            \draw  ##1 -- ++(\pgfkeysvalueof{/pgfplots/box plot width},0pt) |- ##2 -- ++(-\pgfkeysvalueof{/pgfplots/box plot width},0pt) |- ##1 -- cycle;
        },
        /pgfplots/table/.cd,
        y index=\pgfkeysvalueof{/pgfplots/box plot box top index},
        y error expr={
            \thisrowno{\pgfkeysvalueof{/pgfplots/box plot box bottom index}}
            - \thisrowno{\pgfkeysvalueof{/pgfplots/box plot box top index}}
        },
        /pgfplots/box plot
    },
    box plot top whisker/.style={
        /pgfplots/error bars/draw error bar/.code 2 args={%
            \pgfkeysgetvalue{/pgfplots/error bars/error mark}%
            {\pgfplotserrorbarsmark}%
            \pgfkeysgetvalue{/pgfplots/error bars/error mark options}%
            {\pgfplotserrorbarsmarkopts}%
            \path ##1 -- ##2;
        },
        /pgfplots/table/.cd,
        y index=\pgfkeysvalueof{/pgfplots/box plot whisker top index},
        y error expr={
            \thisrowno{\pgfkeysvalueof{/pgfplots/box plot box top index}}
            - \thisrowno{\pgfkeysvalueof{/pgfplots/box plot whisker top index}}
        },
        /pgfplots/box plot
    },
    box plot bottom whisker/.style={
        /pgfplots/error bars/draw error bar/.code 2 args={%
            \pgfkeysgetvalue{/pgfplots/error bars/error mark}%
            {\pgfplotserrorbarsmark}%
            \pgfkeysgetvalue{/pgfplots/error bars/error mark options}%
            {\pgfplotserrorbarsmarkopts}%
            \path ##1 -- ##2;
        },
        /pgfplots/table/.cd,
        y index=\pgfkeysvalueof{/pgfplots/box plot whisker bottom index},
        y error expr={
            \thisrowno{\pgfkeysvalueof{/pgfplots/box plot box bottom index}}
            - \thisrowno{\pgfkeysvalueof{/pgfplots/box plot whisker bottom index}}
        },
        /pgfplots/box plot
    },
    box plot median/.style={
        /pgfplots/box plot,
        /pgfplots/table/y index=\pgfkeysvalueof{/pgfplots/box plot median index},
        semithick,black
    },
    box plot width/.initial=1em,
    box plot x index/.initial=0,
    box plot median index/.initial=1,
    box plot box top index/.initial=2,
    box plot box bottom index/.initial=3,
    box plot whisker top index/.initial=4,
    box plot whisker bottom index/.initial=5,
}

\newcommand{\boxplot}[2][]{
    \addplot [box plot median,#1] table {#2};
    \addplot [forget plot, box plot box,#1] table {#2};
    \addplot [forget plot, box plot top whisker,#1] table {#2};
    \addplot [forget plot, box plot bottom whisker,#1] table {#2};
}

\makeatletter
\IEEEtriggercmd{\fontsize{7.15pt}{9.00pt}\selectfont}
\makeatother
\IEEEtriggeratref{1}

\begin{document}

\title{Distributed Intelligent Illumination Control in the Context of Probabilistic Graphical Models}

\author{M. Cosovic,
        T. Devaja,
        D. Bajovic,
        J. Machaj,
        G. McCutcheon, 
        V. Stankovic,
        L. Stankovic,
        D. Vukobratovic

\thanks{M. Cosovic is with Department for Electric Power Engineering, University of Sarajevo, Bosnia and Herzegovina (e-mail: mirsad.cosovic@gmail.com). T. Devaja, D. Bajovic and D. Vukobratovic are with Department of Power, Electronic and Communications Engineering, University of Novi Sad, Serbia (e-mail: $\{$tijana.devaja, dejanv, dbajovic$\}$@uns.ac.rs). J. Machaj is with Department of Telecommunications and Multimedia, University of Zilina, Slovakia (e-mail: juraj.machaj@fel.uniza.sk). G. McCutcheon is with Ramboll UK, Glasgow, UK (e-mail: Graeme.McCutcheon@ramboll.co.uk). V. Stankovic and L. Stankovic are with Department of Electronic and Electrical Engineering, University of Strathclyde, Glasgow, UK 
(e-mail: $\{$vladimir.stankovic, lina.stankovic$\}$@strath.ac.uk).

This paper has received funding from the European Unions Horizon 2020
research and innovation programme under the Marie Skodowska-Curie grant
agreement No 734331.}}

\markboth{}%
{Shell \MakeLowercase{\textit{et al.}}: Bare Demo of IEEEtran.cls for IEEE Journals}

\maketitle

\begin{abstract}
Lighting systems based on light-emitting diodes (LEDs) possess many benefits over their incandescent counterparts including longer lifespans, lower energy costs, better quality of light and no toxic elements, all without sacrificing consumer satisfaction. Their lifespan is not affected by switching frequency allowing for better illumination control and system efficiency. In this paper, we present a fully distributed energy-saving illumination dimming control strategy for the system of a lighting network which consists of a group of LEDs and user-associated devices. In order to solve the optimization problem, we are using a distributed approach that utilizes factor graphs and the belief propagation algorithm. Using probabilistic graphical models to represent and solve the system model provides for a natural description of the problem structure, where user devices and LED controllers exchange data via line-of-sight communication.
\end{abstract}

\IEEEpeerreviewmaketitle

\section{Introduction}
Lighting systems based on LEDs are becoming the dominant lighting solution due to improved energy efficiency, better quality of light and longer life span. According to the European LED quality charter report, lighting accounts for $15-17\,\%$ of the total energy consumption \cite{euroled}. Similarly, US energy information administration estimated the US residential and commercial sector used around 273 trillion Wh of electricity for lighting in 2017, or about $7\,\%$ of the total electricity consumed \cite{eia}. Globally, there are over 33 billion light sources spending 2650 trillion Wh of electrical energy per year, which is $20\,\%$ of the total global electricity production \cite{humphreys}. 

Replacement of incandescent lamps by LEDs will reduce the lighting consumption to about $5-8\,\%$ of total global consumption. LED technology offers superior control, while LEDs lifespan is not affected by frequent on/off switching, making them suitable for intelligent indoor lighting systems \cite{wang, lee, dong}. Intelligent LED systems usually use dimming feature integrated in a local controller, which can be activated based on local sensor inputs (e.g., daily light intensity).

Different illumination control approaches exist, depending on the system architecture and optimization methods. Authors in \cite{wang, singhvi}, presented the illumination control problem as a tradeoff between energy efficiency and user needs. The utility function is assigned to every user device (UD) with respect to the light intensity. In \cite{lee}, authors present a distributed energy-saving illumination strategy of a network of LEDs and UDs communicating with each other in order to optimize illumination using the message-passing algorithms based on distributed optimization. Authors in \cite{pandharipande} analyzed the energy-efficient LED system in the presence of daylight conditions. In \cite{pan}, the illumination control problem was studied, where users are equipped with portable wireless illumination sensors. Approach where sensors can detect the presence of a user in the office is proposed in \cite{labeodan}, while in \cite{gopalakrishna} authors used advanced machine learning algorithms. In \cite{alice}, a lighting system is described which formulates illumination control as a linear programming problem that aims to both minimize energy usage and meet occupants' preferences. Authors in \cite{koroglu} proposed illumination balancing algorithm which achieves successful control even in the case where decentralized control fails. In \cite{byun}, an intelligent household LED system is described that considers energy efficiency and user satisfaction by utilizing sensors and wireless communication technology.

In this paper, we propose solving the linear optimization problem that trades-off desired level of illumination with the minimal energy consumption in a distributed manner using factor graphs and belief propagation (BP) algorithm. To achieve this goal, we apply the Gaussian BP algorithm, which is recognized as an efficient distributed linear optimization solver with the polynomial-complexity \cite{bickson, mladenov}. Using probabilistic graphical models to represent and solve the system model provides for a very natural description of the problem structure, where LEDs and UDs directly exchange data via point-to-point line-of-sight wireless links. Even if implemented in the centralized framework, it can be flexibly matched to distributed computation resources. The closest to our work is the message-passing solution proposed in \cite{lee}, however, we use the BP-based approach as it is more flexible and converges faster than alternating direction method of multipliers (ADMM).

The paper is organized as follows. After the system model in Sec. II, we provide a sequence of steps in Sec. III and IV, to transform the initial optimization problem to probabilistic graphical model formulation. System-level interpretation of the proposed solution are given in Sec. V, while numerical results are presented in Sec. VI. The paper is concluded in Sec. VII. 

\section{The Dimmable LED Lighting System Model}
\begin{figure}[t]
    \centering
    \includegraphics[keepaspectratio, width=0.40\textwidth]{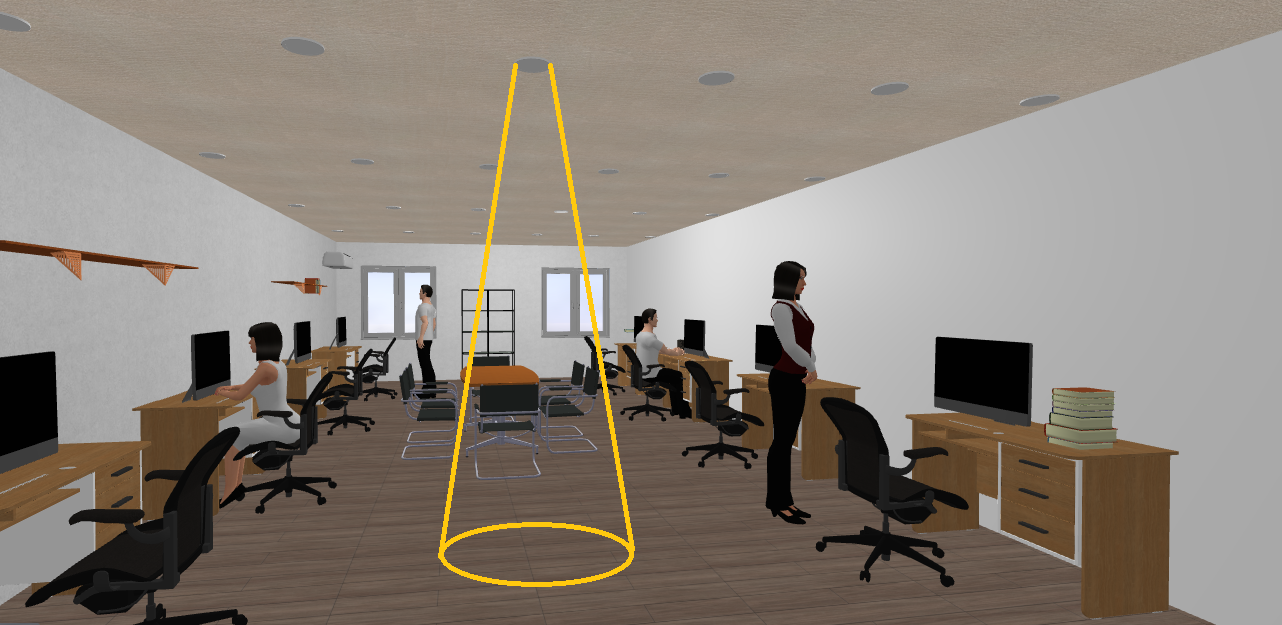}
    \caption{The LED system in the office environment.}   
    \label{fig:office} 
    \end{figure}

We consider a dimmable LED lighting system with $n$ ceiling LEDs associated with dimming vector $\mathbf{y} \in \mathbb{R}^n$. Each LED is equipped with a controller that is able to control the dimming level $y_i$ of the $i$-th LED, adapting its light intensity. The LED illuminates at the maximum light intensity if $y_i=1$ (i.e., LED is set to the maximum power), while $y_i=0$ indicates that the LED is turned off. Parallel to the ceiling is a workspace plane over which spatial illumination is of interest, as shown in Fig. \ref{fig:office}. 
Illuminance distribution $\mathbf{w} \in \mathbb{R}^m$ on the two-dimensional workspace plane, with $m$ user devices with illumination sensors able to measure LED light intensity (i.e., UDs), can be described as follows:
    \begin{equation}
    \begin{aligned}  
    \mathbf{w}=\mathbf{H}\mathbf{y}+\mathbf{p},
    \label{eqn:1}          
    \end{aligned}
    \end{equation}
where $\mathbf{H} \in \mathbb{R}^{m \times n}$ is the illuminance or channel gain matrix, and $h_{ij}$ denotes the illuminance within the zone of the $j$-th UD when $y_i=1$, while all other LEDs are turned off. The vector $\mathbf{p} \in \mathbb{R}^m$ denotes daylight intensity measured by sensors, where $p_j$ represents daylight intensity received at the $j$-th UD zone and it is independent of the LED illumination \cite{wang}. Without loss of generality, in the rest of the paper, we observe the model where only the LED illumination exists, i.e., where $\mathbf{p}=\mathbf{0}$. We also assume that each LED controller has data transfer capabilities via line-of-sight communication with UD, and vice versa,
where the latter is, for convenience, realized in the non-visible (e.g. infrared) part of the optical spectrum. 

In a usual scenario, the desired level of illumination at UDs should be achieved with the minimal energy consumption. Energy consumption of a lighting system can be expressed as a function of the dimming vector \cite{wang}:
    \begin{equation} 
    \begin{aligned}  
    f_0(\mathbf{y})=\frac{\bm{\upepsilon}^T\mathbf{y}+ e_0}{\bm{\upepsilon}^T\mathbf{1} + e_0} = 
    \mathbf{q}^T\mathbf{y} + e,
    \label{eqn:2}          
    \end{aligned}
    \end{equation}
where $\bm{\upepsilon} \in \mathbb{R}^n $ represents the maximum power of LEDs in the system and $e_0$ denotes standby energy consumed by the lighting system. Additionally, we define the normalized power vector $\mathbf{q} \in \mathbb{R}^n$, where $q_i = \epsilon_i/(\bm{\upepsilon}^T\mathbf{1} + e_0)$, while $e = e_0/(\bm{\upepsilon}^T\mathbf{1} + e_0)$. 

The described system leads to the convex optimization problem that includes inequality constraints:
    \begin{equation}
    \begin{aligned}
    & \underset{\mathbf y}{\text{minimize}} & & \mathbf{q}^T\mathbf{y} + e\\
    & \text{subject\;to} & & \mathbf{H}\mathbf{y} \succeq \mathbf{b} \\
    & & & 0  \preceq \mathbf{y} \preceq 1,
    \label{eqn:3}
    \end{aligned}
    \end{equation}
where $\mathbf{b}\in \mathbb{R}^m$ denotes user illuminance requirements. The first constraint represents a relaxation of the target illuminance requirements, allowing the lighting intensity above the required, while the second constraint defines dimming level limits.

\section{The Barrier Method} 
In this section, we utilize the barrier method to solve the linear programming problem \eqref{eqn:3}. We choose the barrier method due to its simplicity over the primal-dual interior-point method. To describe the procedure of finding $\mathbf y$ that minimizes $f_0(\mathbf{y})$ among all $\mathbf y$ that satisfy constraints, it is necessary to reformulate the inequality constrained problem as an equality constrained problem to which Newton’s method can be applied \cite[Sec.~11.2]{boyd}. By introducing the slack variable vector $\mathbf{s} \in \mathbb{R}^{m}$, inequality constraint $\mathbf{H}\mathbf{y} \succeq \mathbf{b}$ turns into the equality constraint $\mathbf{H}\mathbf{y} - \mathbf{s} = \mathbf{b}$. Thus, instead of solving \eqref{eqn:3}, we consider the equivalent problem:
	\begin{equation}
	\begin{aligned}
	& \underset{\mathbf y}{\text{minimize}} & & \mathbf{q}^T\mathbf{y} + e\\
	& \text{subject\;to} & & \mathbf{H}\mathbf{y} - \mathbf{s} = \mathbf{b}\\
	& & & 0 \preceq \mathbf{y} \preceq 1\\ 
	& & & 0 \preceq \mathbf{s} \preceq \infty.
	\label{eqn:4}
	\end{aligned}
	\end{equation}
Using the notation:	
	\begin{equation}
	\begin{aligned}
	{\mathbf{A}} = 
	\begin{bmatrix}
	\mathbf{H}&
	 -\mathbf{I}
	\end{bmatrix}; &&
	{\mathbf{c}} = 
	\begin{bmatrix}
	\mathbf{q}\\
	 \mathbf{0}
	\end{bmatrix}; &&
	{\mathbf{x}} = 
	\begin{bmatrix}
	\mathbf{y}\\
	\mathbf{s}
	\end{bmatrix},
	\end{aligned}
	\label{eqn:5}
	\end{equation}	
where $\mathbf{I}$ is the $m \times m$ identity matrix, and $\mathbf{0}$ is the vector of zeros of dimension $m$, and applying logarithmic barrier function to replace inequality constraints of \eqref{eqn:4}, we obtain:
	\begin{equation}
	\begin{aligned}
	& \underset{\mathbf x}{\text{minimize}} & & t(\mathbf{c}^T\mathbf{x} + e) -  
	\sum_{i=1}^{k} [\ln x_i + \ln (u_i-x_i)]\\
	& \text{subject\;to} & & \mathbf{A}\mathbf{x} = \mathbf{b},
	\label{eqn:6}
	\end{aligned}
	\end{equation}
where $u_i$ denotes the upper bound of the inequalities related with $x_i$, $i = 1, \dots, k$, and $k = n+m$.

To summarize, the optimization problem \eqref{eqn:6} allows for application of Newton's method. However, it is an approximation of the original problem \eqref{eqn:3}, where the quality of the approximation improves as the parameter $t$ grows. Unfortunately, when the parameter $t$ is large, the function is difficult to minimize by Newton’s method, since its Hessian varies rapidly near the boundary of the feasible set \cite[Sec.~11.2.1]{boyd}.

Using Karush-Kuhn-Tucker conditions \cite{aeneas}, it is easy to show that the Newton step $\Delta \mathbf{x}$ and the corresponding dual variable $\mathbf{v}$ satisfy the following system of linear equations:
	\begin{equation}
	\begin{aligned} 
	\begin{bmatrix}
	\mathbf{D} & 
	\mathbf{A}^T\\
	\mathbf{A} & \mathbf{0}
	\end{bmatrix}
	\begin{bmatrix}
	\Delta \mathbf{x}\\
	\mathbf{v}
	\end{bmatrix} =-
	\begin{bmatrix}  
	t\mathbf{c} - \mathbf{d}\\
	\mathbf{Ax} - \mathbf{b}
	\end{bmatrix},
	\end{aligned}
	\label{eqn:7}	
	\end{equation}
where:
	\begin{equation*}
	\begin{aligned} 
	\mathbf{D} &= 
	\mathrm{diag}[x_1^{-2}+(u_1-x_1)^{-2},\dots,x_{k}^{-2}+(u_{k}-x_{k})^{-2}]\\
	\mathbf{d} &= 
	[x_1^{-1}-(u_1-x_1)^{-1},\dots,x_{k}^{-1}-(u_{k}-x_{k})^{-1}]^T. 
	\end{aligned}
	\end{equation*}
The equation \eqref{eqn:7} defines a generalization of the Newton's method taking into account infeasible points. The right hand side contains the block $\mathbf{A}\mathbf{x} - \mathbf{b}$, which is the residual vector for the linear equality constraints, and if $\mathbf x$ is strictly feasible then the residual vanishes (i.e., $\mathbf{Ax} - \mathbf{b} = \mathbf{0}$) \cite[Sec.~10.3]{boyd}. Note that if $\mathbf{y}$ is feasible, then $\mathbf{x}$ is also feasible, because we can take $\mathbf{s} =  \mathbf{H}\mathbf{y} - \mathbf{b}$. Newton's method in each iteration step $\nu$ computes Newton step, and using the line search with a step size $\eta > 0$, updates the state vector:
	\begin{equation}
	\begin{aligned} 
	\mathbf{x}^{(\nu + 1)} = \mathbf{x}^{(\nu)} + \eta \Delta \mathbf{x}^{(\nu)}.
	\end{aligned}
	\label{eqn:8}	
	\end{equation}
The iteration loops are repeated $\nu = \{1,\dots,\nu_{\max} \}$ until the stopping criterion is met, where Newton decrement is commonly used stopping criterion. 

\section{The Barrier Method Using Belief Propagation} 
In this section, we first discuss methods to transform the problem in \eqref{eqn:7} into a linear least-squares (LS) form. Then, we consider restating the resulting LS problem as a maximum-likelihood problem that can be efficiently solved utilizing factor graphs and BP algorithm. 

In general, we can solve the optimization problems \eqref{eqn:7} using a generic method (direct matrix inversion) or a block elimination, where $\mathbf{x}$ can be a feasible or infeasible point\footnote{With a slight abuse of terminology, hereinafter we use terms feasible/infeasible generic method and feasible/infeasible elimination method.}. As an interesting fact, each of the approaches for solving \eqref{eqn:7} will produce a factor graph with different convergence properties. Despite the fact that the block elimination has a much smaller computational cost than the cost of the generic method \cite[Sec.~10.4]{boyd}, we are interested in both methods, due to different convergence properties of the BP algorithm. Next, we observe LS problems of the feasible/infeasible generic and feasible/infeasible block elimination method. 

\begin{itemize}[leftmargin=*]
\item Using the \textbf{infeasible generic method}, we simply obtain a linear LS in the form: 
    \begin{equation}
    \begin{aligned}
    \underset{\Delta \mathbf{x},\mathbf{v} }{\text{minimize}}  \;\;  \Bigg{|\Bigg|} 
    \begin{bmatrix}
    \mathbf{D} & 
    \mathbf{A}^T\\
    \mathbf{A} & \mathbf{0}
    \end{bmatrix}
    {\begin{bmatrix}
    \Delta \mathbf{x}\\
    \mathbf{v}\end{bmatrix}}-
    \begin{bmatrix} 
    \mathbf{d}-t\mathbf{c} \\
    \mathbf{b}-\mathbf{Ax} 
    \end{bmatrix}
    \Bigg|\Bigg|_2^2.
    \end{aligned}
    \label{eqn:9}    
    \end{equation}
     
\item To recall, the \textbf{feasible generic method} reduces to \eqref{eqn:9}, where the block $\mathbf{b} - \mathbf{Ax}$ vanishes.  

\item Using the \textbf{infeasible block elimination method}, we compute the dual variable $\mathbf v =$ $\mathbf v_1$ $-$ $\mathbf v_2$ by solving following LS problems:        
    \begin{equation}
    \begin{aligned}
      &\underset{\mathbf {v}_1}{\text{minimize}} \;\; ||\mathbf{D}^{-1/2}(\mathbf{A}^T\mathbf{v}_1
      -\mathbf{D}\mathbf{x} -\mathbf{d}+t\mathbf{c})
      ||_2^2\\ 
      &\underset{\mathbf {v}_2}{\text{minimize}} \;\; ||\mathbf{A}\mathbf{D}^{-1}\mathbf{A}^T
      \mathbf {v}_2 - \mathbf{b}||_2^2.
    \end{aligned}
    \label{eqn:10}    
    \end{equation}
    
\item A simpler LS form is obtained for the \textbf{feasible block elimination method}:
    \begin{equation}
    \begin{aligned}
      \underset{\mathbf {v}}{\text{minimize}} \;\; ||\mathbf{D}^{-1/2}(\mathbf{A}^T\mathbf{v}+
      t\mathbf{c}-\mathbf{d})||_2^2.
    \end{aligned}
    \label{eqn:11}    
    \end{equation}     
\end{itemize}

Note that once the dual variable $\mathbf{v}$ is determined using the block elimination, the Newton step $\Delta \mathbf{x}$ is obtained as follows: 
     \begin{equation}
    \begin{aligned}
      \Delta \mathbf{x} = 
    \mathbf{D}^{-1}(\mathbf{d}-t\mathbf{c}  
    - \mathbf{A}^T \mathbf{v}). 
    \label{eqn:12}
    \end{aligned}
    \end{equation}    
 
Authors in \cite{bickson, cosovic} show that the linear LS problem can be efficiently solved using Gaussian BP algorithm, and if the algorithm converges, then the fixed point represents a solution of an equivalent LS problem. More precisely, we proposed in \cite{cosovic} how Gaussian BP can be applied as part of the two-level (inner $\tau$ and outer $\nu$) iteration loops for solving non-linear problems akin to Gauss-Newton method, where the inner loop corresponds to the Gaussian BP that solves a linear LS problem.

Let us consider, for the time being, the independent system of linear equations:
    \begin{equation}
    \begin{aligned}
    \mathbf{g} = \mathbf{f}(\mathbf{z}) + \mathbf{r},
    \label{eqn:14}
    \end{aligned}
    \end{equation} 
where $\mathbf{r}$ is an artificial uncorrelated noise with independent and identically distributed entries, and assume that each $r_i$ follows a zero-mean Gaussian distribution with the same variance, e.g., $\mathbf{r} \sim \mathcal{N}(0, \mathbf{I})$. Then, the solution of \eqref{eqn:14} can be found by solving the following LS problem:   
    \begin{equation}
    \begin{aligned} 
    \underset{\mathbf z}{\text{minimize}}  \;\;  ||\mathbf{Fz}-\mathbf{g}||_2^2,
    \label{eqn:13}
    \end{aligned}
    \end{equation} 
where $\mathbf{F}$ is the coefficient matrix for our system \eqref{eqn:14}, with a full column rank. Further, the solution of the LS problem \eqref{eqn:13} can be reformulated as an equivalent maximum likelihood problem that can be solved via maximization of the likelihood function:  
    \begin{equation}
    \begin{gathered}
    \mathcal{L}(\mathbf{g}|\mathbf{z})= 
    \prod_{i} \mathcal{N}(g_i|f_i(\mathbf{z}),\sigma_i^2).
    \end{gathered}
    \label{eqn:15}
    \end{equation}
Due to the fact that any UD is usually illuminated by only a few surrounding LEDs, the function $f_i(\mathbf{z})$ depends on a typically small subset of state variables $\mathbf{z}$. Hence, the likelihood function can be factorized into factors affecting small subsets of state variables. \textbf{This fact motivates solving the problem scalably and efficiently using probabilistic graphical models}. 

From the factorization of the likelihood expression \eqref{eqn:15}, one easily obtains the factor graph. The variables $\mathbf{z}$ determine the set of variable nodes $\mathcal{Z}$, and the set of
factor nodes $\mathcal{F}$ is defined according to likelihood functions $\mathcal{N}(g_i|f_i(\mathbf{z}),\sigma_i^2)$. The factor node connects to the variable node if and only if the variable is an argument of the corresponding function $f_i(\mathbf{z})$. In a nutshell, the structure of the factor graph reflects the structure of the matrix $\mathbf{F}$. More precisely, each row of the matrix $\mathbf{F}$ corresponds to one factor node, while columns, according to the vector $\mathbf{z}$, define variable nodes. A factor node connects to a variable node if and only if the corresponding coefficient of the matrix row is nonzero. Deriving expressions for BP messages exchanged over the factor graph follows similar steps as in \cite{cosovic}. Each message exchanged in Gaussian BP is completely represented using mean and variance. The BP solution of $\mathbf{z}^{(\nu,\tau_{\max})} = \mathbf{z}^{(\nu)}$ in each outer iteration $\nu$ is obtained via the iterative BP algorithm $\tau = \{1,\dots,\tau_{\max}\}$, thereby forming the inner iteration loop. When the BP converged in the one outer iteration loop $\nu$, we update the variables and repeat the process until the stopping criterion is met. 

\section{The Physical Interpretation of Models}
\begin{figure}[ht]
    \centering
    \begin{tikzpicture}
    [every node/.style={node distance=0.6cm},]
    
    \node[name=block, rectangle split, rectangle split parts=4, draw, font=\footnotesize] 
    {\textbf{LED $y_1$}
    \nodepart{second} $h_{11},h_{21},\dots,h_{m1}$ 
    \nodepart{third}  $y_1, \Delta y_1, q_1$
    \nodepart{fourth} $0 \leq y_1 \leq 1$};
    \node (q_dots1) [right=of block] {$\cdots$}; 
    \node[name=block2, rectangle split, rectangle split parts=4, draw, right= of q_dots1,
    font=\footnotesize ] 
    {\textbf{LED $y_n$}
    \nodepart{second} $h_{1n},h_{2n},\dots,h_{mn}$ 
    \nodepart{third}  $y_n, \Delta y_n, q_n$
    \nodepart{fourth} $0 \leq y_n \leq 1$};
    
    \node[name=block3, rectangle split, rectangle split parts=4, draw, below= of block, 
    font=\footnotesize] 
    {\textbf{UD $w_1$}
    \nodepart{second} $h_{11},h_{12},\dots,h_{1n}$ 
    \nodepart{third}  $s_1, \Delta s_1, v_1, b_1$
    \nodepart{fourth} $0 \leq s_1 \leq \infty$};    
    \node (q_dots2) [right=of block3] {$\cdots$};
    \node[name=block4, rectangle split, rectangle split parts=4, draw, right= of q_dots2,
    font=\footnotesize ] 
    {\textbf{UD $w_m$}
    \nodepart{second} $h_{m1},h_{m2},\dots,h_{mn}$ 
    \nodepart{third}  $s_m, \Delta s_m, v_m, b_m$
    \nodepart{fourth} $0 \leq s_m \leq \infty$};    
    
    \path[draw,-latex,>=stealth, <->] (block) -- (block3);
    \path[draw,-latex,>=stealth, <->] (q_dots1) -- (q_dots2);    
    \path[draw,-latex,>=stealth, <->] (block) -- (q_dots2);
    \path[draw,-latex,>=stealth, <->] (block2) -- (q_dots2);
    \path[draw,-latex,>=stealth, <->] (block2) -- (block4);
    \end{tikzpicture}
    \caption{Data structure and communication patterns between LEDs and UDs relative to the 
    physical and optimization models.}
    \label{fig:parameters}
    \end{figure}
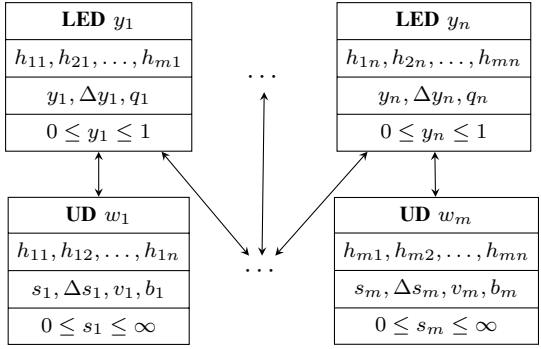

Providing a physical interpretation assumes starting from the initial system \eqref{eqn:1} and optimization problems \eqref{eqn:3} and \eqref{eqn:4} that are in turn transformed into \eqref{eqn:7} and \eqref{eqn:8}. We assume LEDs and UDs exchange data via line-of-sight optical wireless communications, as shown in Fig. \ref{fig:parameters}. Parameters associated with LEDs and UDs are mostly defined according to the physical system. The $i$-the LED controller contains information about the $i$-th column of the channel gain matrix $\mathbf{H}$, the normalized power $q_i$, dimming level $y_i$, with associated Newton step $ \Delta y_i$ (artificial parameter), and dimming limits $0 \leq y_i \leq 1$. The $j$-th UD is associated with the $j$-th row of the channel gain matrix $\mathbf{H}$ and user illuminance requirement $b_j$. Additionally, artificial parameters that UD contains are information related to the slack and dual variables $s_j$, $\Delta s_j$, $0 \leq s_j \leq \infty$ and $v_j$.

It can be noted that physical models \eqref{eqn:1} and \eqref{eqn:3} were transformed and extended to get LS problems \eqref{eqn:9} - \eqref{eqn:11}, along with \eqref{eqn:12}. For this purpose, we give a detailed analysis of factor graphs that can be constructed according to different LS problems. Essentially, the resulting factor graph, through $\mathbf{F}$ and $\mathbf{g}$, should preserve and sustain communication and data structure associated with LEDs and UDs. More precisely, each factor node must be defined based on recipes given in Fig. \ref{fig:parameters}. 

The generic method \eqref{eqn:9} produces the same matrix $\mathbf{F}$ for the infeasible and feasible problem. Consequently, the core structure of the factor graph will be the same, with variable nodes $\mathcal{Z} = \{\Delta \mathcal{Y},\Delta \mathcal{S},\mathcal{V} \}$ and three different sets of factor nodes $\mathcal{F} = \{\mathcal{F}_{u}, \mathcal{F}_{lu},\mathcal{F}_{ul} \}$: 
\begin{enumerate}[leftmargin=*]
\item The factor node $f_{u,j} \in \mathcal{F}_{u}$ is defined locally according to the corresponding UD parameters, and connects variable nodes $v_j \in \mathcal{V}$ and $\Delta s_j \in \Delta \mathcal{S}$.
\item The factor node $f_{lu,i} \in \mathcal{F}_{lu}$ is defined locally according to the corresponding LED parameters, and connects the variable node $\Delta {y}_i \in \Delta \mathcal{Y}$ with the group of variable nodes $\mathcal{V}_i$ $\subseteq$ $\mathcal{V}$, reflecting line-of-sight illumination and communication structure between the LED and the set of neighboring UDs.
\item The factor node $f_{ul,j} \in \mathcal{F}_{ul}$ connects the variable node $\Delta s_j \in \Delta \mathcal{S}$ with the group of variable nodes $\Delta \mathcal{Y}_j \subseteq \Delta \mathcal{Y}$, reflecting the line-of-sight illumination/communication relationship between the UD and the set of surrounding LEDs. For the feasible problem, factor node can be defined locally according to the corresponding UD, while for the infeasible problem, additional communication overhead is required, where before each outer iteration loop $\nu$, LEDs need to send $\mathcal{Y}_j \subseteq \mathcal{Y}$ to the corresponding UD. Note that $\mathcal{Y}$ is defined by the vector $\mathbf{y}$.
\end{enumerate}

To summarize, the feasible/infeasible generic method preserves and sustains the structure given in Fig. \ref{fig:parameters} and allows for application of fully distributed message-passing algorithms according to the physical structure of the problem.  

\begin{example}[The probabilistic model for the feasible/infeasible generic method] In this toy example, we observe two LEDs $y_1$ and $y_2$, where both LEDs communicate with UDs $w_1$ and $w_2$. The set of factor nodes consists of $\mathcal{F}_{u} =$ $\{f_{u,1},$ $f_{u,2}\}$, $\mathcal{F}_{lu} =$ $\{f_{lu,1},$ $f_{lu,2}\}$ and $\mathcal{F}_{ul} =$ $\{f_{ul,1},$ $f_{ul,2}\}$.
    \begin{figure}[ht]
    \centering
    \includegraphics[width=7.0cm]{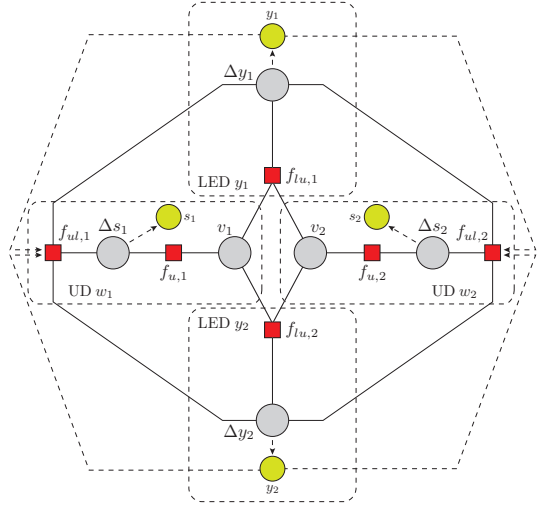}
    \caption{The example of probabilistic model for the feasible/infeasible generic method.}
    \label{fig:graph_generic}
    \end{figure}  
          
The output of the inner iteration loop provides the values of variable nodes $\mathcal{Z} = \{\Delta y_1, \Delta y_2, \Delta s_1, \Delta s_2, v_1, v_2\}$. Then, using \eqref{eqn:8}, we update variables $\{y_1,y_2,s_1,s_2 \}$ and repeat the process until the stopping criterion is met. Finally, after the algorithm converges, we obtain optimal values of dimming levels $\{y_1,y_2\}$. The communication overhead for the feasible problem contains only BP messages, while the infeasible problem additionally requires sending values of dimming levels $\{y_1,y_2\}$ to the corresponding UDs for the purpose of defining factor nodes $\mathcal{F}_{ul}$ in each outer iteration $\nu$.     
\end{example}

The infeasible block elimination method produces two LS problems \eqref{eqn:10}, causing the existence of two disconnected factor graphs. The first factor graph, that reflects the structure of the matrix $\mathbf{F} = \mathbf{D}^{-1/2}\mathbf{A}^T$, preserves the pattern structure given in Fig. \ref{fig:parameters}. In contrast, the second factor graph, described by $\mathbf{F} = \mathbf{A}\mathbf{D}^{-1}\mathbf{A}^T$, compromises the physical pattern of data transfer between LEDs or UDs. Consequently, we cannot represent the infeasible block elimination method via suitable factor graph representation, thus we do not consider it in this paper. 

The feasible block elimination method produces the factor graph according to $\mathbf{F} = \mathbf{D}^{-1/2}\mathbf{A}^T$ and $\mathbf{g} = \mathbf{D}^{-1/2}(t\mathbf{c}-\mathbf{d})$, with variable nodes $\mathcal{Z} = \mathcal{V}$, and two different sets of factor nodes $\mathcal{F} = \{\mathcal{F}_{u}, \mathcal{F}_{lu}\}$:     
\begin{enumerate}[leftmargin=*]
\item The factor node $f_{u,j} \in \mathcal{F}_{u}$ is defined locally according to the corresponding UD parameters, and connects variable node $v_j \in \mathcal{V}$.
\item The factor node $f_{lu,i} \in \mathcal{F}_{lu}$ is defined locally according to the corresponding LED parameters, and connects group of variable nodes $\mathcal{V}_i \subseteq \mathcal{V}$, following the line-of-sight illumination/communication relationship between the LED and neighbouring UDs.
\end{enumerate}    
This model along with \eqref{eqn:12} also preserves and sustains structure given in Fig. \ref{fig:parameters}. 

\begin{example}[The probabilistic model for the feasible block elimination method] In this toy example, we observe three LEDs $y_1,$ $y_2$ and $y_3$, where UDs $w_1$ and $w_2$ communicate with $y_1,y_2$ and $y_2,y_3$, respectively. The set of factor node consists of $\mathcal{F}_{u} =$ $\{f_{u,1},$ $f_{u,2}\}$ and $\mathcal{F}_{lu} =$ $\{f_{lu,1},$ $f_{lu,2},$ $f_{lu,3}\}$.      
    \begin{figure}[ht]
    \centering
    \includegraphics[width=6.3cm]{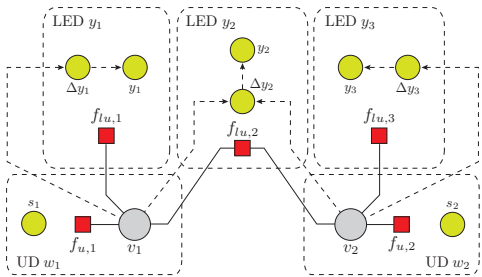}
    \caption{The example of probabilistic model for the feasible block elimination method.}
    \label{fig:graph_elimination}
    \end{figure} 

The output of the inner iteration loop provides values of variable nodes $\mathcal{Z} = \{v_1, v_2\}$. After that, each UD sends the value of the corresponding variable node to the set of neighboring LEDs. Finally, using \eqref{eqn:12} and \eqref{eqn:8} each LED and UD compute locally the values of variables $\{y_1,y_2,y_3,s_1,s_2 \}$ and repeat the process until the stopping criterion is met, when we obtain optimal values of dimming levels $\{y_1,y_2,y_3\}$.       
\end{example}
    
In comparison with the generic method, the block elimination has a much simpler structure, with a smaller number of variable and factor nodes.

\vspace{-2.0mm}
\section{Numerical Results}
In all simulated models, we are employing a loopy Gaussian BP algorithm using the synchronous scheduling with randomized damping \cite{cosovic}. The synchronous scheduling updates messages from variable to factor nodes, and messages from factor nodes to variable nodes, in parallel in respective half-iterations. Additionally, we damped each mean value message from a factor node to a variable node independently with predefined probability, evaluating the mean value message as a linear combination of the message from the previous and the current iteration.

\textbf{Convergence:} We consider the feasible/infeasible generic and feasible elimination method, with the goal of investigating the convergence of the BP algorithm. The methods are tested using an open-plan office space with a square-shaped floor area of side length $15\,\text{m}$ and height $3\,\text{m}$, where LED sources are fixed on an equidistant grid on a ceiling plane, providing for $n = 100$ LEDs. We generate 200 random configurations, where we randomly distribute $m = 15$ UDs on the workspace plane, in order to obtain average convergence performances. 

The convergence of the BP algorithm depends of the spectral radius $\rho$ of the matrix that governs evolution of means from factor nodes to variable nodes, and the BP converges to a unique fixed point if and only if $\rho < 1$. In our case, the BP will converge if all spectral radii for each outer iteration $\rho^{(\nu)} < 1$, $\nu = 1,\dots,\nu_{\max}$. Consequently, as we shown in \cite{cosovic}, the BP converges to a unique fixed point if and only if $\rho_{\max}<1$, where:
        \begin{equation}
        \begin{aligned}
          \rho_{\max} = 
        \max\{\rho^{(\nu)}:
        \nu= 1,\dots,\nu_{\max}\}. 
        \end{aligned}
        \label{eqn:16} 
        \end{equation}

\begin{figure}[ht]
    \centering
    \begin{tikzpicture}
      \begin{axis}[width=7.5cm, height=5.5cm,
       x tick label style={/pgf/number format/.cd,
       set thousands separator={},fixed},
       xlabel={Spectral Radius $\rho_{\max}$},
       ylabel={Empirical CDF $F(\rho_{\max})$},
       label style={font=\footnotesize},
       grid=major,
       legend style={at={(0.05,0.8)},anchor=west,font=\scriptsize},
       legend cell align={left},
    legend columns=1,         
       ymin = 0, ymax = 1.1,
       xmin = 0.55, xmax = 1.25,
       xtick={0.6,0.7,0.8,0.9,1,1.1,1.2},
       tick label style={font=\footnotesize},
       ytick={0,0.1,0.2,0.3,0.4,0.5,0.6,0.7,0.8,0.9,1.0}]
  
    \addplot[mark=square*,mark repeat=58, mark size=1.4pt, black]
       table [x={x}, y={y}] {./figure/plot1/feasible_elimination.txt};
       \addlegendentry{Feasible Elimination} 
  
    \addplot[mark=diamond*,mark repeat=70, mark size=1.5pt, red] 
       table [x={x}, y={y}] {./figure/plot1/feasible_generic.txt};
       \addlegendentry{Feasible Generic}
       
    \addplot[mark=otimes*, mark repeat=68, mark size=1.5pt, blue]
    table [x={x}, y={y}] {./figure/plot1/infeasible_generic.txt};       
       \addlegendentry{Infeasible Generic}
       
      \end{axis}
    \end{tikzpicture}
    \caption{The maximum spectral radii $\rho_{\max}$ over outer iterations $\nu$ 
    for the feasible elimination and feasible/infeasible generic method.}
    \label{fig:spectral_radii}
    \end{figure}
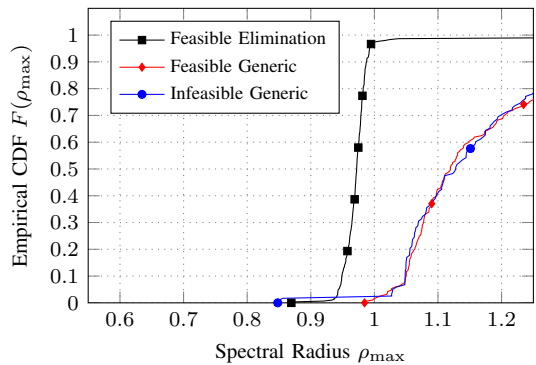

Fig. \ref{fig:spectral_radii} shows empirical cumulative density function (CDF) $F(\rho_{\max})$ of spectral radius $\rho_{\max}$ for observed method. The feasible elimination method is superior in terms of the spectral radius, where we record convergence with probability $0.97$. In contrast, the feasible/infeasible generic method converged with the negligible probability of $0.01$. To summarize, we identify the BP-based feasible elimination method as a good candidate for efficient and distributed solver in the context of the intelligent illumination control.                    

\textbf{Communication Overhead:} The BP messages exchanged between LEDs and UDs represent the communication overhead. In a single inner iteration, the BP-based feasible elimination method exchanges messages from a variable node to a factor node, i.e, messages from UDs to LEDs, and vice versa. In the following, we are interested in assessing the number of inner iterations $\tau$ per outer iteration $\nu$, and corresponding time to complete one outer iteration $\nu$. Hence, the BP algorithm in the inner iteration loop is running until the following criterion is reached:    
        \begin{equation}
        \begin{gathered}      
        |\mathbf{z}_{f \to v}^{(\nu,\tau)}-
        \mathbf{z}_{f \to v}^{(\nu, \tau-1)}| \preceq 10^{-14}
        \;\;\mathrm{or}\;\;
        \tau_{\max} = 2000,
        \end{gathered}
        \label{num_break2}
        \end{equation}
where $\mathbf{z}_{f \to v}$ represents the vector of mean-value messages from factor nodes to variable nodes. We set the convergence criteria very conservative with the aim of achieving high level of accuracy, although for the practical applications, the illumination system is does not require high precision of dimming states (e.g., $10$-bit dimming control with $1024$ dimming levels is considered highly precise). Further, assuming that the processing time is negligible compared to the message transmission, it is possible to give a rough time estimate of duration of a single outer iteration $\nu$. Our study focuses on low-rate visible-light communication system with $250\,\text{kbit/s}$ that performs exchange of BP messages between LEDs and UDs, where each BP message represents a $64\,\text{bit}$ data block. We note that one can also consider low-rate RF technologies such as IEEE 802.15.4 based ZigBee. Our motivation here is a scenario where UDs are static and fixed at office space desk, in line-of-sight conditions with respect to ceiling LEDs, and are able to detect occupant presence, measure illuminance and perform simple BP processing and message exchange with neighboring LEDs. Thus we close the whole system, both illumination and communication, using LED-based technologies.

With this aim, we observe a large open-plan office space with square floor area of side $50\,\text{m}$ and height $3\,\text{m}$. In the first scenario, we fixed number of LEDs to $n = 625$ ($25 \times 25$), and consider $m =$ $\{50,$ $60,$ $70,$ $80,$ $90,$ $100\}$ UDs. Next, we fixed number of UDs to $m = 50$, and observe systems with $n =$ $\{625,$ $676,$ $729,$ $784,$ $841,$ $900\}$ LEDs. Additionally, for each pair of numbers LEDs-UDs, we randomly distributed $m$ UDs over $200$ configurations. 

\begin{figure}[ht]
    \captionsetup[subfigure]{oneside,margin={1.6cm,0cm}}
    \begin{tabular}{@{}c@{}}
    \subfloat[]{\label{plot3a}
    \centering
    \begin{tikzpicture}
    \begin{axis} [box plot width=1.0mm,
    y tick label style={/pgf/number format/.cd,
    set thousands separator={},fixed},
    xlabel={Number of UDs $m$},
    ylabel={Inner itarations $\tau$},
    grid=major,           
    xmin=0, xmax=7, ymin=-100, ymax =2100,    
    xtick={1,2,3,4,5,6},
    xticklabels={50,60,70,80,90,100},
    width=8cm,height=4.5cm,
    tick label style={font=\footnotesize}, label style={font=\footnotesize},
    legend style={draw=black,fill=white,legend cell align=left,font=\tiny,
    legend pos=south west}]
    \boxplot [
    forget plot, fill=blue!30,
    box plot whisker bottom index=1,
    box plot whisker top index=5,
    box plot box bottom index=2,
    box plot box top index=4,
    box plot median index=3] {./figure/plot2/n_625_box.txt};   

    \addplot[only marks, mark options={draw=black, fill=blue!30},mark size=0.2pt] 
    table[x index=0, y index=1] {./figure/plot2/n_625_outliers.txt};      

    \end{axis}
    \end{tikzpicture}}
    \end{tabular}\\
    \begin{tabular}{@{}c@{}}
    \subfloat[]{\label{plot3b}
    \begin{tikzpicture}
    \begin{axis} [box plot width=1.0mm,
    y tick label style={/pgf/number format/.cd,
    set thousands separator={},fixed},
    xlabel={Number of LEDs $n$},
    ylabel={Inner itarations $\tau$},
    grid=major,           
    xmin=0, xmax=7, ymin=-100, ymax =2100,    
    xtick={1,2,3,4,5,6},
    xticklabels={625,676,729,784,841,900},
    width=8cm,height=4.5cm,
    tick label style={font=\footnotesize}, label style={font=\footnotesize},
    legend style={draw=black,fill=white,legend cell align=left,font=\tiny,
    legend pos=south west}]
    \boxplot [
    forget plot, fill=blue!30,
    box plot whisker bottom index=1,
    box plot whisker top index=5,
    box plot box bottom index=2,
    box plot box top index=4,
    box plot median index=3] {./figure/plot2/m_50_box.txt};   

    \addplot[only marks, mark options={draw=black, fill=blue!30},mark size=0.2pt] 
    table[x index=0, y index=1] {./figure/plot2/m_50_outliers.txt};    
    
    \end{axis}
    \end{tikzpicture}}
    \end{tabular}
    \caption{The number of inner iterations $\tau$ for the system with 
    $m =$ $\{50,$ $60,$ $70,$ $80,$ $90,$ $100\}$ UDs and $n = 625$ 
    LEDs (subfigure a), and the system with 
    $n =$ $\{625,$ $676,$ $729,$ $784,$ $841,$ $900\}$ LEDs and $m = 50$ UDs (subfigure b).}
    \label{fig:iteration}
\end{figure}
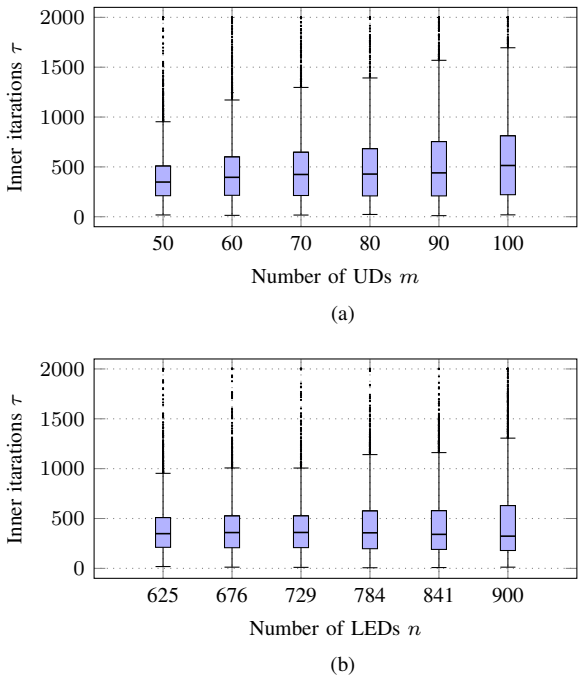

Fig. \ref{fig:iteration} shows distribution of the total number of inner iterations $\tau$ over outer iterations $\nu$ and $200$ configurations. It can be concluded that the number of inner iterations $\tau$ growing more in the case of increasing the number of UDs, compared to the case of an increasing number of LEDs. To roughly estimate the time to broadcast all BP messages in one outer iteration $\nu$, we consider median values. Median values for the first scenario are bounded between $348$ and $550$ inner iterations, regarding time to broadcast messages can be estimated as $(64/250) \cdot$ $348=$ $89.1\,\text{ms}$, and $(64/250) \cdot$ $514=$ $131.6\,\text{ms}$. Median values of the second scenario are around $350$ iterations, with corresponding time to broadcast all messages in one outer iteration $\nu$ as $(64/250) \cdot$  $350=$ $89.6\,\text{ms}$. Hence, the BP-based algorithm can provide the solution in a reasonable time, for observing models. 

\section{Conclusions} 
We presented scalable and efficient distributed approach to solve the illuminance optimization problem using the BP algorithm. The aim of the work was to set a ground for BP-based framework for smart illumination. In our future work, we plan to investigate BP-based version of interior-point methods, which, based on our initial results, seems to be even more efficient for BP-based implementation. We also plan to explore the solution of the optimization problem when the channel gain matrix is unknown. More precisely, authors in \cite{gligoric} presented an approach based on compressed sensing, where the UD is able to easily recover the channel gain matrix. The presented model is suitable for integration in the BP framework, which allows the algorithm to estimates the channel gain matrix and minimize the energy consumption according to the desired illumination level simultaneously in the time continuous process.

\bibliographystyle{IEEEtran}
\bibliography{cite}

\end{document}